\documentclass[twocolumn]{elsartwide}

\usepackage{graphics}
\usepackage{graphicx}
\usepackage{epsfig}

\usepackage{amssymb}

\begin{document}

\begin{frontmatter}



\title{A liquid xenon ionization chamber in an all-fluoropolymer vessel}


\begin{center}

\large{EXO Collaboration} \\
\end{center}
\author[stanford]{F.~LePort}, 
\author[stanford]{A.~Pocar\corauthref{cor}}, 
\corauth[cor]{Corresponding author. Address: Physics Department, Stanford University, 382 Via Pueblo Mall, Stanford, CA  94305, USA.  Tel: +1-650-725-2342; fax: +1-650-725-6544.} 
\ead{pocar@stanford.edu}  
\author[stanford]{L.~Bartoszek}, 
\author[stanford]{R.~DeVoe},
\author[stanford]{P.~Fierlinger}, 
\author[stanford]{B.~Flatt}, 
\author[stanford]{G.~Gratta}, 
\author[stanford]{M.~Green}, 
\author[stanford]{T.~Koffas\thanksref{thomas}},
\thanks[thomas]{Now at CERN, Geneva, Switzerland}
\author[stanford]{M.~Montero Diez},
\author[stanford]{R.~Neilson}, 
\author[stanford]{K.~O'Sullivan},
\author[stanford]{S.~Waldman\thanksref{sam}},
\thanks[sam]{Now at Caltech, Pasadena CA, USA}
\author[stanford]{J.~Wodin}, 
\author[apt]{D.~Woisard},
\author[neuchatel]{E.~Baussan}, 
\author[slac]{M.~Breidenbach},
\author[slac]{R.~Conley},
\author[csu]{W.~Fairbank Jr.}, 
\author[laurentian]{J.~Farine}
\author[slac]{C.~Hall\thanksref{carter}}, 
\thanks[carter]{Now at University of Maryland, College Park MD, USA}
\author[csu]{K.~Hall}, 
\author[laurentian]{D.~Hallman},
\author[carleton]{C.~Hargrove},
\author[slac]{J.~Hodgson},
\author[csu]{S.~Jeng},
\author[bama]{D.S.~Leonard}, 
\author[slac]{D.~Mackay},
\author[neuchatel]{Y.~Martin},
\author[slac]{A.~Odian}, 
\author[neuchatel]{L.~Ounalli}, 
\author[bama]{A.~Piepke}, 
\author[slac]{C.Y.~Prescott}, 
\author[slac]{P.C.~Rowson}, 
\author[slac]{K.~Skarpaas}, 
\author[neuchatel]{D.~Schenker}, 
\author[carleton]{D.~Sinclair},
\author[itep]{V.~Stekhanov}, 
\author[carleton]{V.~Strickland}, 
\author[laurentian]{C.~Virtue},
\author[neuchatel]{J.-L.~Vuilleumier},
\author[neuchatel]{J.-M.~Vuilleumier}, 
\author[slac]{K.~Wamba}, 
\author[neuchatel]{P.~Weber}

\address[stanford]{Physics Department, Stanford University, Stanford CA, USA}
\address[apt]{Applied Plastics Technology, Inc., Bristol, RI, USA}
\address[neuchatel]{Institut de Physique, Universit\'e de Neuchatel, Neuchatel, Switzerland}
\address[slac]{Stanford Linear Accelerator Center, Menlo Park CA, USA}
\address[csu]{Physics Department, Colorado State University, Fort Collins CO, USA}
\address[laurentian]{Physics Department, Laurentian University, Sudbury ON, Canada}
\address[carleton]{Physics Department, Carleton Univerisity, Ottawa ON, Canada}
\address[bama]{Department of Physics and Astronomy, University of Alabama, Tuscaloosa AL, USA}
\address[itep]{Institute for Theoretical and Experimental Physics, Moscow, Russia}

\begin{abstract}
A novel technique has been developed to build vessels for liquid xenon ionization detectors
entirely out of ultra-clean fluoropolymer.    We describe the advantages in terms of low radioactivity 
contamination, provide some details of the construction techniques, and show the 
energy resolution achieved with a prototype all-fluoropolymer ionization detector.
\end{abstract}

\begin{keyword}
PTFE \sep fluoropolymer  \sep double beta decay \sep xenon \sep EXO \sep low background

\PACS  23.40.-s \sep 26.65.+j \sep 29.40.Mc \sep 81.05.Lg
\end{keyword}
\end{frontmatter}

\section{Introduction}
\label{s:intro}

Plastics have been studied and used as structural materials for particle detectors 
requiring ultra low levels of radioactive contamination for over a decade by 
experiments such as Chooz~\cite{Chooz}, Palo Verde~\cite{PaloVerde}, SNO~\cite{SNO}, 
Borexino/CTF~\cite{Borexino}, KamLAND~\cite{KamLAND}, and MUNU~\cite{MUNU}. 
In these detectors, acrylic, nylon, and EVOH\footnote{Ethyl vinyl alcohol derivative film} 
were used for the containers of the innermost, active liquid volumes. 
Because of their light weight, low atomic mass, and production processes that 
involve efficient chemical separation stages, selected plastics have generally shown 
very low levels of long-lived radioactivity, particularly in the naturally occurring isotopes 
$^{40}$K, $^{232}$Th, and $^{238}$U. 
Polycarbonate has also been used for cryogenic, hermetic vessels, most 
notably in small bubble chambers~\cite{Lexan}, because of its mechanical stability, strength, and
compatibility with low temperatures.    Unfortunately all polycarbonate samples
(both in the form of raw pellets and molded plates) measured by our 
group~\cite{EXO_Rad_measurements} have shown substantial radioactive contaminations
(especially $^{40}$K), 
disqualifying them for use in low background experiments.

EXO (``Enriched Xenon Observatory'') is a program~\cite{EXO} aimed at building a 
ton-class double beta decay detector with xenon-136. The plan is to use enriched xenon 
(80\% $^{136}$Xe) as source and detection medium.
While the EXO collaboration is planning to build a ton-scale detector with the ability
to retrieve and identify the Ba atom produced in the $\beta\beta$ decay 
of $^{136}$Xe, an intermediate scale (200~kg of enriched xenon) detector without the Ba 
tagging feature, called ``EXO-200'', is currently under advanced construction~\cite{exo-200}.
The Xe in EXO-200 (and, possibly, in the full size EXO) is kept in liquid phase
(LXe) at a temperature around 170~K at a pressure of 1~atm.    The two 
electrons produced in the $\beta\beta$ decay are detected in an electric field of 
1 to 4~kV/cm by the simultaneous readout of ionization and scintillation.     This 
technique has been shown to provide superior energy resolution~\cite{Conti}, 
as required to suppress backgrounds without sharp energy features, such as the $2\nu \beta\beta$ decay and $\gamma$-ray Compton tails.  
In EXO-200 the VUV scintillation light (175~nm) is detected by ``bare'' (i.e. without their standard encasing) large area avalanche photodiodes (LAAPDs)\footnote{SD155-9718 - 16~mm Deep UV LAAPD from Advanced Photonix Inc., Camarillo CA, USA.}, while the ionization signal is collected by crossed grids, hence measuring the
total energy of the decay and its three dimensional location. The third dimension
is provided by the drift time using the scintillation signal as a ``start''.
The active LXe has the shape of a cylinder about 40~cm long and 40~cm in diameter.
The vessel with the LXe is submerged in HFE-7000\footnote{Novec{\textregistered} Engineered Fluid HFE-7000 
(C$_3$F$_7$OCH$_3$) is a fluorinated heat transfer fluid by the 3M Company, St. Paul MN, USA.} contained in a low activity, copper cryostat.   The HFE-7000 is an ultra-clean fluid that is 
in liquid phase in a broad range of temperatures, encompassing 300~K and 170~K.
This fluid is used as the innermost $\gamma$ and neutron shield and thermal bath 
to maintain a uniform temperature around the LXe vessel.   The presence of
a large thermal mass of HFE-7000 fluid makes it possible to use xenon containers with poor thermal conductivity such as plastics or thin metallic shells.

The end-point energy for the $0\nu\beta\beta$ decay in $^{136}$Xe 
(2457.8~keV~\cite{endpoint}) is substantially higher than that of most, but not all, 
radioactive decays in the U and Th decay chains.  In addition, $^{40}$K and other
less common backgrounds are also relevant for a clean reconstruction of the 
broad energy distribution originating from the $2\nu\beta\beta$ decays.  
The high monetary value of the enriched Xe makes it impractical to use it for shielding the detection volume from background events originating from the xenon vessel material.
The intrinsic background requirements for all 
construction components and, in particular, the relatively heavy LXe vessel are therefore 
very challenging to achieve.    Additional constraints are given by the cryogenic 
temperatures and the LXe purity requirement with respect to electronegative 
contaminations, which would reduce the electron lifetime in the detector.

In this paper we will discuss the use of a particular  
fluoropolymer for the 
construction of a large cryogenic vessel containing the LXe.   We will also 
present data from an ionization chamber built entirely out of such fluoropolymer.   Structural
performance and details of the construction technology used for the chamber will be
the subject of a future publication.  Although copper was ultimately chosen for the first generation EXO-200 chamber for scheduling reasons, 
the developments of a fluoropolymer chamber described here may be used at a future stage of EXO and by other groups.

\section{Fluoropolymer as a vessel material}

Fluoropolymers such as polytetrafluoroethylene (PTFE) have been used in the past in LXe detectors as a VUV reflector 
and volume displacer~\cite{VUV_reflector_displacer}.    It is therefore known that it is 
compatible with the clean environment required to drift electrons over large distances
of LXe.   In addition, fluoropolymer parts are generally known to retain their structural 
integrity at cryogenic temperatures.     DuPont Teflon{\textregistered}\footnote{Teflon{\textregistered} is a registered trademark of E.I. du Pont de Nemours \& Company.} TE-6472~\cite{TE6472} is a variety 						
of {\it modified} PTFE developed for use in the semiconductor industry and hence produced 
with high purity standards.     Blanks are sintered by first pressing fine pellets in a
mold and then baking the material at a specific oven cycle.     Levels of radioactive contaminants 
in the raw pellets of TE-6472 have been measured~\cite{EXO_Rad_measurements}
with neutron activation analysis 
(NAA) and were found to be $(1.65\pm 0.17) \times 10^{-9}$~g/g for K (1$\sigma$),  
$< 0.26 \times 10^{-12}$~g/g for Th and $< 0.35\times 10^{-12}$~g/g for U, at 95\%~CL.
NAA results for several other elements confirm the extremely low level of impurities 
in the material.  It was found that other batches of the same product have a similar level
of purity and that the process of sintering, once appropriate quality control procedures are 
implemented, does not appreciably degrade the purity of the material.  Final parts are 
obtained by conventional machining followed by an etch and rinse that remove possible surface 
contaminations left by the cutting tools.    Most high grades of fluoropolymers from other suppliers were found not to reach the purity levels observed in TE-6472, which are among the very 
best ever measured in solid materials.   Detailed comparison of radioactive contaminants in 
several plastics and many other materials will be reported in the near future~\cite{EXO_Rad_measurements}.

An important property of the TE-6472 fluoropolymer is the addition of a thermoplastic fluoropolymer, or 
{\it modifier}, to the raw PTFE in order to turn it into a weldable material.  In addition, the pelletized
form of the raw material results in a more homogeneous finished material
than conventional PTFE parts.   
The specific gravity of finished TE-6472 is 2.2~g/cm$^3$ (at 20~$^{\circ}$C). 
The welding of finished parts together is carried out by heating
the material while applying substantial pressure.  Weldability is essential for our
purpose since it makes it possible to seal the vessel hermetically, without the use 
of flanges, gaskets, or fasteners.   Substantial amounts of material can thus be 
removed from locations close to the fiducial volume of the detector and the
difficulty of sealing the detector using materials with vastly different coefficients of thermal 
expansion, (in this case modified PTFE\footnote{The thermal expansion coefficient of PTFE is $2\times 10^{-4}$~K$^{-1}$, much larger than that of metals.}  and metals), is eliminated.    The possibility of fusing parts together also allows for the construction of 
large components from smaller elements.     
This is important considering the increasing difficulty of compression molding large parts in one piece, a process that requires a high tonnage press with substantial platen-to-platen clearance and ram travel to accommodate the typical 3:1 powder compression ratio.

\section{An all-fluoropolymer detector vessel}

\begin{figure} 
\begin{center}
\includegraphics[width=8.5cm]{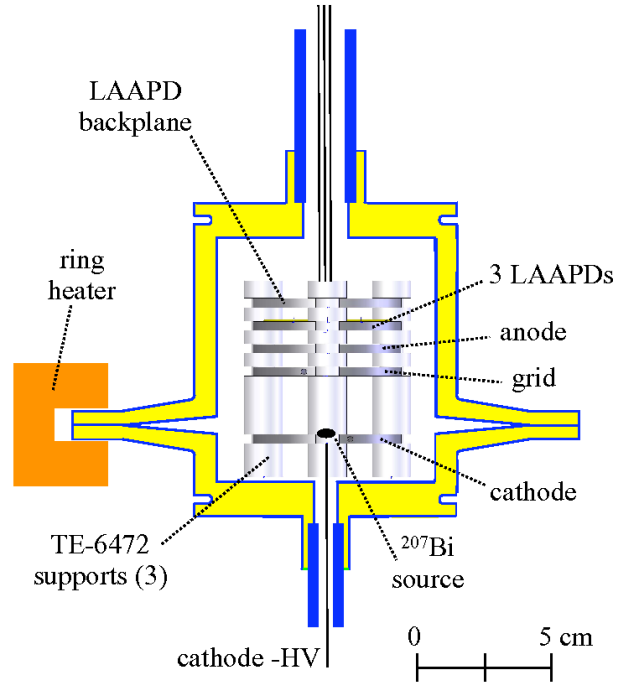}
\vskip 0.1cm
\caption{
Schematic drawing of the all-fluoropolymer LXe vessel with the welding fixture
schematically shown on the left.   The upper and lower halves are machined to their final shapes,
including the weld flanges with their thermal expansion-compliant regions.  
A different type of ``socket'' weld is used to couple straight pipe sections to
the ports on the top and bottom of the vessel.
TE-6472 supports, independent from the vessel construction, are used to 
support and space the electrodes in the stack and the LAAPD package. The drift distance 
(between cathode and grid) is 2.5 cm. All stainless steel electrode planes and
rings are 3 mm thick. A $^{207}$Bi electron capture source is
plated at the center of the cathode plane (visible in fig.~\ref{fig:stack}). }
\label{fig:vessel_design}
\end{center}
\end{figure}

\begin{figure}[htb!!]
\begin{center}
\includegraphics[width=7.8cm]{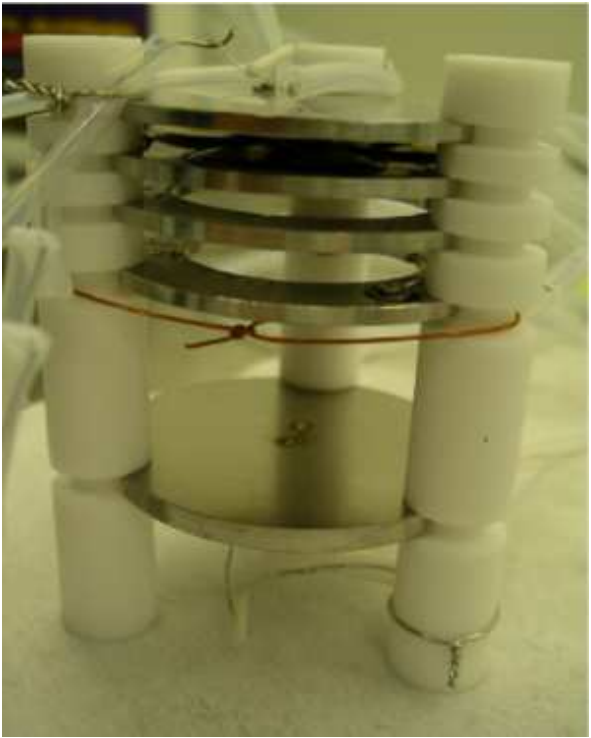}
\vskip 0.1cm
\caption{Photograph of the electrode stack and LAAPD package.    Supports independent 
on the chamber vessel are built using the same fluoropolymer. Note the $^{207}$Bi source in the center of the cathode (bottom) plate. RTDs can also be seen above the top plate and below the cathode.}
\label{fig:stack}
\end{center}
\end{figure}

In the course of this program several types of all-fluoropolymer LXe time projection chamber (TPC) prototype vessels were built. 
The sketch of one of them is shown in Figure~\ref{fig:vessel_design}. The vessel is 
comprised of two halves which are welded together to form a cylindrical cell of 
7.9~cm inner diameter and 8.8~cm inner length.   The wall thickness is 0.8~cm. 
Two long pipes made out of a similar\footnote{Dyneon{\textregistered}
modified PTFE TFR-1105 by the 3M Company, St. Paul MN, USA.} type of of 
modified PTFE as that of the 
vessel are welded to special socket features at both ends. These 
commercially available pipes are 2.5~cm and 1.9~cm in outer diameter respectively, 
with a wall thickness of 3~mm. They have similar welding properties but were not 
screened for radioactivity. The ability to make long pipes from TE-6472 was 
developed for later prototypes.
Two pipes (inlet and outlet) are included in order to have inline recirculation
and purification of the xenon because of the expected outgassing of electronegative 
impurities from the fluoropolymer walls of the vessels.
As discussed below, it was later found that Xe purification is not necessary.   
Moreover, having access to the detector via two 
pipes allows one to route the cable for the high voltage cathode to one feedthrough
and the ones for lower voltage bias and signal connections at the anode end of the detector to a separate one (see section~\ref{s:results}).

\begin{figure}[b!!]
\begin{center}
\includegraphics[width=8cm]{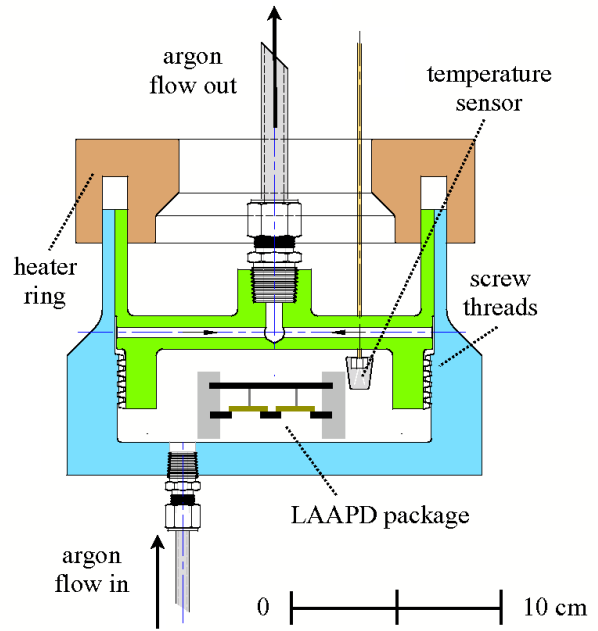} 
\vskip 0.1cm
\caption{
Cross section of a fluoropolymer chamber designed to isolate and purge the welding fumes from the LAAPDs.    Arrows indicate the path of a 6~l/min Ar purge active during the weld and the cool-down period.
Note that in this particular version of a vessel the field weld is achieved by radial (as opposed
to longitudinal) compression. For simplicity, threaded fittings not suitable for a cooldown were used for this test of welding compatibility with LAAPDs.}
\label{fig:APDs}
\end{center}
\end{figure}

The vessel's construction involves several different welds, that are performed in a specialized 
shop using ovens and localized heater bands and plates.   Each half-body is machined from 
billets which are obtained by fusing smaller molded pre-forms in an oven in order to reach the
required size. 
Socket welds are then performed to connect pipes that then transition to the 
metal piping of the xenon system, as described below.
The detector instrumentation (ionization cell with LAAPDs and temperature sensors, 
described in section~\ref{s:results} and shown in figures~\ref{fig:vessel_design} and~\ref{fig:stack}) 
is then inserted into the vessel that is subsequently closed with a field weld using a hot 
clamp.  In the vessel shown in figure~\ref{fig:vessel_design} the field weld is made using 
a disk-like feature designed to have enough compliance to accommodate rather large 
thermal distortions during the weld.  
The temperature inside the fluoropolymer cell is monitored with resistance temperature detectors 
(RTDs) during the welding cycle and kept below 150~$^\circ$C. 
Specific problems were observed with a drastic reduction in the quantum 
efficiency of the LAAPDs at 175~nm, 
after they had been sealed inside the welded fluoropolymer chambers.
Although a substantial array of high sensitivity 
surface analyses failed to show fluorine contamination on the LAAPDs active
face, a specific geometry of the field weld (fig.~\ref{fig:APDs}), designed with 
localized water cooling and able
to efficiently flush gases produced in the welding process and to isolate the 
LAAPDs from the welded region, completely eliminates the problem. We note that 
the bare LAAPDs have been found to be, by far, the components most sensitive 
to welding fumes.    Thin wire grids and meshes made of different metals appear
to be substantially less sensitive to this issue.

\begin{figure}[htb!!]
\begin{center}
\includegraphics[width=8cm]{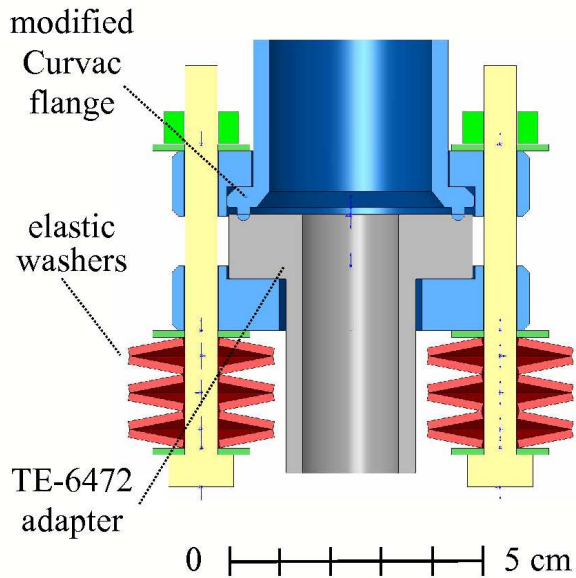} 
\vskip 0.1cm
\caption{
Schematic drawing of the cryogenic fluoropolymer-to-metal seal.   The modified stainless
steel Curvac flange (see text) is pressed against a fluoropolymer mating flat with 
constant force.}
\label{fig:curvac}
\end{center}
\end{figure}

Because of the relatively large permeability of fluoropolymers to He at room 
temperature, leak checking of the finished vessels with He leak detectors requires 
some care.    Experience has shown, however, that the onset of the signal from 
diffusion, for fluoropolymers of any reasonable thickness, is much slower than that of a 
real leak, so that a clear discrimination between the two phenomena is easily 
achieved.     It was generally found that the fluoropolymer vessels with their 
welds and fluoropolymer plumbing are very reliable and He leak tight even after several 
cycles from room temperature to 170~K.  Typical leak rates 
$< 10^{-4}$~cm$^3$ mbar s$^{-1}$ are observed~\cite{teflon-paper}.

Cryogenic fluoropolymer-to-metal transitions are required in order to interface the 
all-fluoropolymer vessel to more conventional plumbing.    Reliable transitions for
$\sim 2.5$~cm diameter pipes have been obtained by pressing a modified
Curvac flange\footnote{Curvac is a trade mark of ULTEK Corp (Palo Alto, CA)
and it refers to a vacuum flange similar in size to the 2-3/4-inch Conflat,
but with a round profile sealing edge instead of a knife edge.} onto a carefully
machined flat surface of the TE-6472 material, as schematically shown in 
figure~\ref{fig:curvac}.   Using a linear pressure of 82~kg/cm, the circular 
cross section of the Curvac feature penetrates the fluoropolymer up to about half of
its radius.   A stack of elastic washers insures a constant force against the 
seal as materials contract or expand and, over long periods of time, 
creep (in the case of fluoropolymers).   Note that, as temperature changes, the curvac feature ``glides''
radially on the fluoropolymer surface, without apparent loss of the seal.    Transitions
built with this technique have been shown to work reliably between 380 and 
170~K even after several temperature cycles.   Although the transitions are
rather bulky, where radioactive backgrounds are a concern they can be located
far away from the active region of the detector, by providing long fluoropolymer tubes
to move them behind shielding material (HFE-7000 in the EXO-200 design).

\begin{figure}[t!!]
\begin{center}
\includegraphics[width=7.8cm]{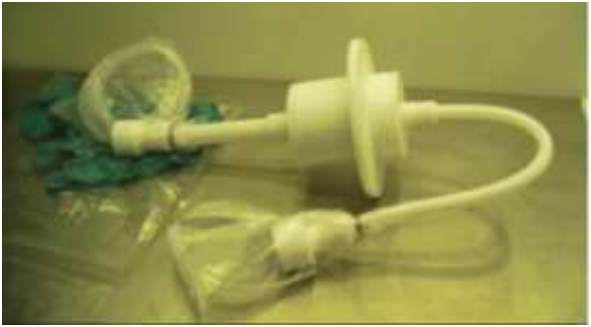} 
\vskip 0.1cm
\caption{
Photograph of a finished chamber of the type illustrated in fig.~\ref{fig:vessel_design}.   
The two pipes are used for 
xenon recirculation and as conduits guiding the detector wiring to the
outside.   Electrical feedthroughs are, in this prototype, conventional 
(metal-ceramics), mounted after the fluoropolymer-to-metal transition.   The high
potential ($\sim 5$~kV) required for the cathode is fed through the 
pipe opposite to the one feeding the anode, grids and APD potentials and
signals.}
\label{fig:finished_chamber}
\end{center}
\end{figure}

A chamber built with the present technique is shown in 
fig.~\ref{fig:finished_chamber}.   The fluoropolymer parts of the fluoropolymer-to-metal 
transitions are visible at the ends of the pipes.    Electrical wires, insulated 
with PFA\footnote{Perfluoroalcoxy (PFA) polymers are thermoplastic perfluorinated polymers often used for extruded tubes and sheets.} 
jackets are also visible (in plastic bags to 
maintain cleanliness).     The tube with the ``U'' bend is located at the bottom
of the cryostat containing the cold HFE-7000 and the two fluoropolymer-to-metal transitions 
are mounted near the top flange of the cryostat.    All wiring goes through the transitions 
and onto conventional metal-ceramic feedthroughs branching off the xenon feed lines.
We note that because of the poor heat conduction of fluoropolymers, a large chamber built with this 
technique would require an external Xe condenser.

\section{Permeation of HFE-7000 in fluoropolymers}

Permeation of HFE-7000 through the fluoropolymer walls of the chambers was thoroughly investigated because of the possible implications in terms of electronegative contamination of the LXe.
Long term permeation measurements were undertaken in two fluoropolymers 
(DuPont PFA-440HP and 3M modified PTFE TFM-1700)  before TE-6472 fluoropolymer was selected as the most suitable candidate material for a LXe vessel.

\begin{figure}[t!!]
\begin{center}
\includegraphics[width=8.1cm]{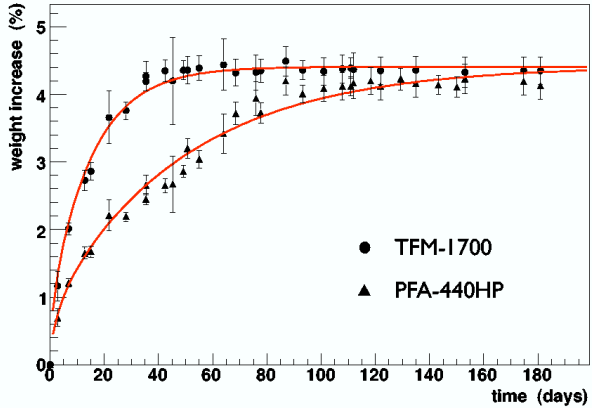} 
\vskip 0.1cm
\caption{
Time evolution of HFE-7000 permeation into TFM-1700 (a modified PTFE similar to DuPont TE-6472) and PFA-440HP. The data were taken with a thin sample and are fit assuming diffusion into an infinite, parallel plate slab (see text).}
\label{fig:hfe-intake}
\end{center}
\end{figure}

Dog-bone-shaped samples, 1.6 mm-thick, 38 mm long, and 15 mm wide were immersed in HFE-7000 at room temperature and their weight recorded as a function of time, until saturation was reached. Data for TFM-1700 and PFA-440HP are shown in fig.~\ref{fig:hfe-intake}: full saturation is reached after 1.5 and 5 months respectively. Data were analyzed assuming a purely diffusive intake into a slab with infinite, parallel walls.  This assumes that HFE-7000 intake through the edges of the sample, which constitute $\sim 15\%$ of the exposed surface, is negligible.  The effect of such approximation is to overestimate the permeation rate, yielding a slightly conservative result.  The solution for the fractional weight gain versus time $\lambda(t)$ is (see e.g.~\cite{boas} for a derivation):
$$
\lambda (t) =\; S\,\left [1 -\frac{4}{\pi^2}\sum _{n=1}^{\infty} \frac{1-(-1)^n}{n^2} e^{-\left (\frac{n\pi}{L} \right )^2 Dt}  \right ],
\label{eq:lambda}
$$
where $S$ is the solubility parameter (i.e. the saturation amount of HFE-7000 expressed as fraction of the initial weight of the sample), $L$ the thickness of the sample, and $D$ the diffusion coefficient.
A fit to the data yields, for TFM-1700, $S=(4.4\pm0.2)\%$ and $D=(1.9\pm 0.7)\times 10^{-9}$~cm$^2$~s$^{-1}$, while, for PFA-440HP, $S=(4.4\pm 0.5)\%$, $D=(5.9\pm 2.1)\times 10^{-10}$~cm$^2$~s$^{-1}$.

A second test of the permeation properties of HFE-7000 through fluoropolymers was performed by exposing one side of a thin PFA-440HP membrane to HFE-7000 at room temperature while pumping out the other side. The evacuated volume was sampled with a residual gas analyzer measuring characteristic HFE-7000 mass peaks (e.g. 69, 81, 90, 119, 131 amu). A value $D=3\times 10^{-10}$~cm$^2$~s$^{-1}$ was obtained by measuring the time needed for the 81 amu concentration to reach 95\% of the saturation level.
This value is within a factor a 2 of the value obtained by sample immersion ($S$ cannot be extracted from this type of measurement).

Samples of TFM-1700 and PFA-440HP were also tested, with limited accuracy, for permeation of two more heat transfer fluids, FC-87 and HFE-7100\footnote{Fluorinert{\textregistered} FC-87 (C$_5$F$_{12}$) and Novec{\textregistered} Engineered Fluid HFE-7100 (C$_4$F$_9$OCH$_3$) are products of the 3M Company, St. Paul MN, USA.}, that had also been considered for use in EXO-200.  It was found that, for both plastics, $S\sim5\%$ for HFE-7100 and $S\sim9\%$ for FC-87. $D$ ranges between $\sim3\times 10^{-10}$~cm$^2$~s$^{-1}$ for HFE-7100 in PFA-440HP and $\sim 2\times 10^{-9}$~cm$^2$~s$^{-1}$ for FC-87 in TFM-1700, consistent with PFA-440HP being less permeable than TFM-1700.
Permeation tests were also performed on TFM-1700 samples cooled down to -10~$^{\circ}$C.  While saturation had not yet been reached at the time of writing, data suggest that $S$ is larger by possibly $\sim 50\%$, while the diffusion coefficient is more than ten times smaller than at room temperature.
TE-6472 is a modified PTFE much like TFM-1700, and the permeation properties of the two are expected to be similar.  Short term tests were performed in order to compare TE-6472 and TFM-1700, showing a $\sim10\%$ lower value for TE-6472.

\begin{figure}[htb!!]
\begin{center}
\includegraphics[width=8.3cm]{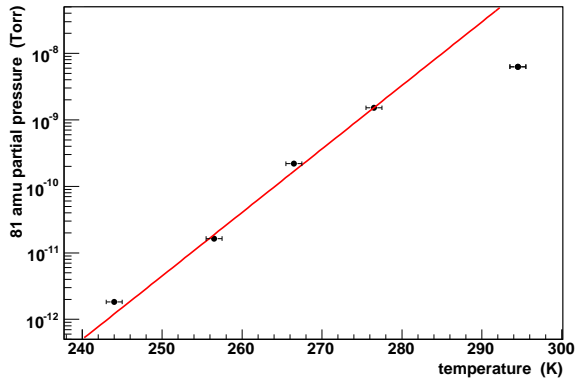} 
\vskip 0.1cm
\caption{
Temperature dependance of HFE-7000 permeation in TE-6472. Permeation falls exponentially as the temperature decreases, with a 10-fold reduction every 10.5~K.  The data point point at room temperature was not included in the fit. }
\label{fig:permeation-vs-temp}
\end{center}
\end{figure}

As a final, crucial test, a sample of TE-6472 soaked in HFE-7000 was inserted in a LXe cell, operated as a xenon purity monitor. The cell, similar to the one described in~\cite{xpm}, has a long drift region, which allows it to measure electron drift times up to 4 ms. After the sample was inserted into it, the cell was evacuated and cooled down prior to xenon liquefaction. The intensity of the HFE-7000 mass peaks was measured versus temperature, providing information on the HFE-7000 emanation rate. 
Below 247~K, no HFE-7000 signal could be detected on top of the RGA background ($\sim$$10^{-12}$ Torr partial pressure for mass 81 amu). The data above 247~K show a 10-fold reduction in the permeation rate for every 10.5~K decrease in temperature, as shown in fig.~\ref{fig:permeation-vs-temp}.  An extrapolation to 170~K indicates an extremely small release into the LXe.  Indeed, no specific sign of impurity specifically attributable to the HFE-7000 was observed when measuring electron lifetime.						

The intake of fluorinated fluids causes fluoropolymers to swell and deteriorates their mechanical properties. Our study shows that TFM-1700 and PFA-440HP linearly expand $\sim$$3$\% when saturated with FC-87 or HFE-7100.  A two-fold reduction of their bulk modulus and yield strength is also observed; such degradation is largest for TFM-1700 in FC-87 and smallest for PFA-440HP in HFE-7100.

\section{Operation of an all-fluoropolymer liquid xenon ionization chamber}
\label{s:results}

As illustrated in figures~\ref{fig:vessel_design} and~\ref{fig:stack} , the ionization chamber
consists of an LAAPD package and 3 horizontal stainless steel structures stacked vertically, 
submersed in 0.5~liters of LXe.    Ionization electrons are created in the volume between 
the lowest plane (cathode) and the grid directly above it and are 
collected by an anode behind the grid.    The cathode consists of a solid 
stainless steel plate, whereas the grid and anode are made of electroformed 
nickel meshes with 90\% optical transparency, mounted on stainless steel rings.     
Good electron transparency is obtained by maintaining the anode-to-grid field at twice the
value of the drift field. 
Three LAAPDs, located above the anode plane, detect scintillation 
photons.   However in the measurements shown here, the quality of the signals
from the LAAPDs was rather poor because of the damage caused by the early 
welding process.  Therefore LAAPD signals were only used for triggering and 
timing purposes and not for energy reconstruction.     Later welding tests,
that preserved the full functionality of LAAPDs, were performed, for simplicity,
in setups with no grid structure.

After seal-welding the vessel and connecting it to the Xe feed and recovery systems,
the entire system is baked at 380~K under vacuum for several days.   During this 
operation the interior of the cryostat is purged with Ar so that no air or water
vapor can diffuse into the walls of the fluoropolymer vessel and, eventually, contaminate
the xenon.   After the system is returned to
room temperature, the Ar in the cryostat is replaced with HFE-7000
and cooling is started.     Commercial xenon is purified by passing it through an 
Oxisorb{\textregistered} cartridge\footnote{Oxisorb{\textregistered}  cartridge from Messer-Griesheim Gmbh, 47805 Krefeld, Germany.} eight times.   While hot zirconium getters are typically
used for low background experiments, Oxisorb{\textregistered}, known to introduce radon contamination in the xenon, 
was adequate for this test.   	
All the data shown here was taken between 165.5 and 171~K
and between 1200 and 1470~Pa.

\begin{figure}[t!!]
\begin{center}
\includegraphics[width=8.1cm]{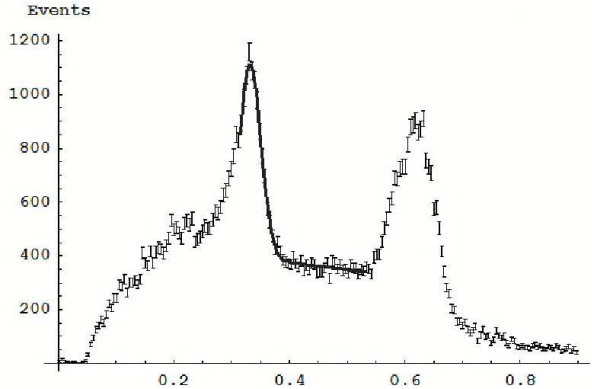} 
\vskip 0.1cm
\caption{Ionization energy spectrum collected at a drift field of 1.5~kV/cm. The two 
peaks correspond to the two internal conversion e$^-$ lines from the $^{207}$Bi 
source, at 570 keV and 1064 keV (K-shell electron events at 975 keV are also visible as a shoulder to the 1064 keV peak). 
In order to evaluate the energy resolution at 570 keV,
the data is fit to a Gaussian plus a straight line as shown. The higher energy peak is not included in the fit. The horizontal axis 
shows the amplitude of the pulse in arbitrary units.}
\label{fig:spectrum}
\end{center}							
\end{figure}

Ionization in the LXe is produced by both the 570~keV and 1064~keV $\gamma$ ray lines 
following the $\beta$ decay of $^{207}$Bi as well as by internal conversion electrons.    
 The source is prepared as a monoatomic 
layer of radioactive Bi electroplated onto a 25~mm$^2$ piece of electroformed nickel 
mesh, which is then spot-welded to the cathode plate. The data presented here 
were collected at drift fields of 0.25, 0.5, 0.75, 1, 
and 1.5~kV/cm.   Events are selected with drift time corresponding to the cathode-to-grid distance to obtain a relatively pure sample of internal conversion (IC) electrons which all interact in very close proximity to the source.
The electron signal at the anode was detected by a low noise, charge-sensitive 
Amptek A250 preamplifier, shaped with
a Canberra model 2020 spectroscopy amplifier, sampled with a transient digitizer,
 and recorded.
Special care is required to reduce pick-up noise.  An all-fluoropolymer vessel,
with no intrinsic metallic shield around the electrodes. is an
unconventional layout for an ionization detector.  Proper grounding and shielding
must be provided by other means (for instance, by the cryostat enclosure). 

In the off-line analysis charge signals are fit to the following 5-parameter
model of the drift to the anode of a Gaussian ionization distribution:
$$
A\, \sigma \left[ {\rm erf}\left(\frac{t_0}{\sqrt{2}\sigma} \right) - {\rm erf} \left(\frac{-t+t_0}{\sqrt{2}\sigma} \right)  \right] e^{\frac{-t+t_0}{\tau}} + B.
$$
Here $A$ is the amplitude, $t_0$ the effective drift-time, $\sigma$ the rise-time parameter, 
$\tau$ the integrator fall-time, and $B$ a vertical offset.  The readout was 
calibrated in units of electron charge by injecting a signal into the pre-amplifier 
through a 1 pF calibrated capacitor. The electronic noise of the pre-amplifier was found to follow 
a Gaussian distribution with standard deviation of 597 electrons.    Data acquisition is 
triggered when a delayed coincidence between scintillation and ionization signals 
is observed.   The delay accounts for the drift velocity of electrons and varies 
with electric field.   The selection of IC events, which occur along the main axis of the ionization cell,
alleviates the effects of possible non-uniformities in the electric field across the active volume
of the detector, which is not equipped with field-shaping electrodes.  

\begin{figure}[t!!]
\begin{center}
\includegraphics[width=8cm]{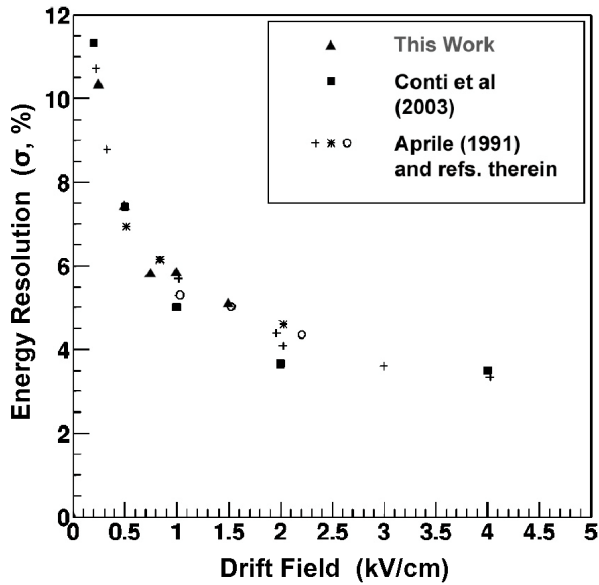} 
\vskip 0.1cm
\caption{Ionization energy resolution for 570~keV electrons in LXe at several drift fields. 
The triangles represent the resolution observed in this experiment. The other 
symbols represent the resolution for ionization obtained by other 
experiments~\cite{resolution_digest,Conti} using chambers of conventional construction.}
\label{fig:resolution_digest}
\end{center}
\end{figure}

An ionization spectrum obtained at 1.5~kV/cm is shown in figure~\ref{fig:spectrum}. 
The 570~keV peak of the $^{207}$Bi spectrum is fit to a Gaussian plus a straight line as 
shown.  The noise-subtracted energy resolution is shown as a function of 
the drift field in figure~\ref{fig:resolution_digest}.  Such resolution for 570 keV is found to be $\sigma/E = 5.1\%$ at 1.5~kV/cm, in good
quantitative agreement with the results obtained by our group and others using conventional construction techniques~\cite{resolution_digest}.

The chamber described here was run for 10~days without re-purification of the LXe. 
No degradation of the signal amplitude or of the resolution was observed.

\section{Conclusions}

Fluoropolymer construction of vessels for TPCs using LXe and other liquefied noble gases
has been demonstrated for the first time.    Because of the very low radioactive
contaminations in the special fluoropolymer variety discussed here, this exotic technique has great 
promise for low background experiments.

\section{Acknowledgments} 

We greatly thank S.~Ebnesajjad, S.~Libert, and D.N.~Washburn of the DuPont Company for 
many invaluable discussions and for providing samples and data on fluoropolymers.   
We would like to thank Applied Plastics Technology, Inc. for their PTFE compression molding, welding development efforts, continuing manufacturing support, and for the enthusiasm shown toward the project.  
We are very greatful to the DuPont Company for supplying TE-6472 material free of
charge to the experiment.   We acknowledge Kenji Kingsford and the Advanced Plastic Division of the Saint-Gobain Company for providing samples and performing pull tests on many of them. 
We are particularly grateful to the Waste Isolation Pilot Plant (WIPP) for their enthusiastic and enduring support.
This work has been supported, in part, by US DoE Grant DE-FG03-90ER40569-A019 and US DoE Grant DE-FG29-02AL68086.  APT's vessel fabrication and welding process development has been supported by US DoE-SBIR Grant DE-FG02-04ER83903.



\begin{thebibliography}{00}
\bibitem{EXO_Rad_measurements} A paper reporting on the extensive trace radioactivity measurements performed in EXO will be published in the near future.
\end{thebibliography}
\end{document}